# Magnetoresistance of a spin MOSFET with ferromagnetic MnAs source and drain contacts

Ryosho. NAKANE[1,a)], Tomoyuki HARADA[1], Kuniaki SUGIURA[1], Satoshi SUGAHARA[2], and Masaaki TANAKA[1,b)]

[1]*Deptartment of Electrical Engineering and Information Systems, The University of Tokyo, 7-3-1 Hongo, Bunkyo-ku, 113-8656 Tokyo, Japan*

[2]*Imaging Science and Engineering Laboratory and Dept. of Electronics and Applied Physics, Tokyo Institute of Technology, 4259-G2-14 Nagatsuta, Yokohama, Kanagawa 226-8502, Japan*

Spin-dependent transport was investigated in a spin metal-oxide-semiconductor field-effect transistors (spin MOSFET) with ferromagnetic MnAs source and drain (S/D) contacts. The spin MOSFET of bottom-gate type was fabricated by photolithography using an epitaxial MnAs film grown on a silicon-on-insulator (SOI) substrate. In-plane magnetoresistance showed spin-valve-type hysteretic behavior, when the measurements were performed with constant source-drain and source-gate biases. By comparing with the magnetization-related resistance change resulting from the MnAs contacts, we conclude that the spin-polarized electrons are injected from the MnAs source into the Si MOS inversion channel, and detected by the MnAs drain.

Recent-extensive investigations on semiconductor-based spintronic devices are primarily aiming at creating key devices beyond the complementary metal-oxide-semiconductor (CMOS) scaling.[1-8] For this purpose, spintronic devices using spin-polarized carriers in Si have generated much attention and some demonstrations have begun to reveal physical properties in such systems.[4-8] Among Si-based spintronic devices, recently-proposed spin metal-oxide-semiconductor field-effect transistors (spin MOSFETs)[9-11] are expected to have potential advantages, such as high spin-dependent effects, high active device performance, and their good compatibility with the CMOS technology. The analysis of the potential performance is mandatory for integrating spin MOSFETs with a CMOS platform, and also experimental demonstration and understanding of output characteristics are strongly needed. Although it was recently reported that spin transport in bulk Si is robust,[4-8] there is no experimental study on the transport in Si-MOS structures. In this paper, we present the spin-dependent transport in a MOSFET with ferromagnetic MnAs source/drain (S/D) contacts.

To clarify the spin-dependent transport in lateral structures comprising ferromagnetic metal 1 (FM1)/semiconductor(SC)/ferromagnetic metal 2(FM2), two types of measurement geometries, which are referred to as the local geometry and nonlocal geometry, have been used. In the nonlocal geometry, a voltage difference between the FM1(or FM2) contact and the SC channel is measured while a constant current flows from the FM2(or FM1) contact to the SC channel. An advantage of the nonlocal geometry over local geometry is that important parameters in spin-dependent transport, such as spin diffusion length and spin lifetime, can be easily estimated since the charge current will not contribute to the output voltage change. On the other hand, in the local geometry, a voltage difference between the FM1 and FM2 contacts is measured while a constant current flows from the FM1 (FM2) contact to the FM2 (FM1) contact via the SC channel. In this geometry, magnetoresistance contains a lot of magnetization-dependent phenomena, such as anisotropic magnetoresistance (AMR), local Hall effect (LHE),[12] and spin-valve effect. Since AMR and LHE are induced from one FM contact, AMR and LHE signals can also be detected in FM/SC/nonmagnetic metal (NM) structures.[12] Thus, we can quantitatively extract the spin-valve effect by comparing the magnetoresistance for a FM/SC/FM structure with that for a FM/SC/NM structure. When this evaluation is used, both the device structure and the *I-V* characteristics must be almost identical in the FM/SC/FM and FM/SC/NM structures, because AMR and LHE also depend on both the structure and the applied bias. Despite the difficulties mentioned above, we emphasize here that characterizing the spin-valve effect in the local geometry is indispensable for spintronic device applications since a large output current (or voltage) with a large resistance change is strongly needed for the integration with other electronic devices.

Ferromagnetic MnAs films with hexagonal crystal structure can be epitaxally grown on Si(001) substrates using molecular beam epitaxy (MBE).[13] In the previous reports, we have shown that MnAs films have a low Schottoky barrier height with Si(001) for electrons,[14] and that MOSFETs with MnAs S/D contacts shows good transistor characteristics.[15] In this letter, we study the spin-dependent transport of a spin MOSFET with MnAs S/D contacts (referred to as MnAs MOSFET) fabricated on a p-type (001) silicon-on-insulator (SOI) substrate, where the buried oxide (BOX) layer and the bottom Si substrate in the SOI substrate were used as a gate dielectric and a gate electrode, respectively. The main device-structure parameters in this study are as follows: The channel SOI (Si) thickness is 20 nm, the BOX thickness is 200 nm, the S/D area is 100×100 μm$^2$, the physical channel width $W_G$ is 110 μm, and the physical channel length $L_G$ defined by the gap between S/D is 10 μm. The fabrication process is basically the same as those in the previous report.[15] Notable modifications from the previous report are as follows; (1) the thickness of the MnAs S/D layers is different from each other (20 nm and 100 nm) to employ the difference in the coercivity between the MnAs S/D layers, and (2) the SOI thickness is thinned to 12 nm using a $H_2SO_4$ and $H_2O_2$ mixture to apply the gate voltage effectively. Figure 1(a) and (b) show the schematic cross-section and the optical micrograph of a fabricated MOSFET device. In order to measure the magnetization-related resistance change resulting from the MnAs contact itself, such as AMR and LHE, MOSFETs with MnAs source (S) and nonmagnetic AuSb (with 5 % Sb in weight) drain (D) contacts (referred to as asymmetric MOSFETs) were prepared from the same MBE-grown substrate, where the ohmic-like contact between AuSb and the Si conduction band was formed using ramp annealing at 350ºC in a $N_2$ atmosphere. For another reference, n-type MOSFETs (referred to as n-MOSFETs) with the same device structure were also fabricated by replacing both of the MnAs S/D contacts with nonmagnetic phosphorous-doped Si areas.

Figure 2(a) shows the source-drain current $I_{DS}$ vs. source-drain bias voltage $V_{DS}$ output characteristics measured at 3 K for the MnAs MOSFET, where the source−gate bias $V_{GS}$ varied from 0 to 100 V in steps of 20 V. Despite the low Schottoky barrier height between MnAs and the Si conduction band, non-linear exponential $I_{DS} - V_{DS}$ characteristics are seen at low $V_{DS}$ and $V_{GS}$ biases. This means that the tunnel emission transport through the reverse-biased Schottky barrier between the source and the channel dominates the current conduction at 3 K. The $I_{DS} - V_{DS}$ characteristics for the MnAs MOSFET are almost identical to those for an asymmetric MOSFET with the same $L_G$. On the other hand, the source-drain current $I_{DS}$ vs. source-gate bias voltage $V_{GS}$ output characteristics were also measured with a constant $V_{DS}$ = 50 mV for the MnAs MOSFET and a n-MOSFET with the same $L_G$ = 10 μm, as shown in Fig. 2(b). The difference in those two characteristics indicates the influence of the MnAs/Si Schottky barrier on the characteristics; in the MnAs MOSFET, the subthreshold region shifts by ~20 V to the higher value, and the on-current above 20 V decreases by two orders of magnitude. Using other n-MOSFETs with $L_G$ = 20, 40, 60, 100 μm, the parasitic resistance was estimated to be a few tens of ohms, indicating that the n-MOSFET resistance is almost the same as the channel resistance $R_{CH}$. This result allows us to quantitatively estimate the total Schottoky barrier resistance $R_{SB}$ in the MnAs MOSFET, which consists of the source-channel and drain-channel junction resistances. Typical values of the estimated $R_{SB}$ and $R_{CH}$ ($L_G$ = 10 μm) are as follows: $R_{SB}$ = 215 kΩ and $R_{CH}$ = 614 Ω for $V_{DS}$ = 50 mV and $V_{GS}$ = 20 V, $R_{SB}$ = 15 kΩ and $R_{CH}$ = 270 Ω for $V_{DS}$ = 50 mV and $V_{GS}$ = 50 V, $R_{SB}$ = 180 Ω and $R_{CH}$ = 18 Ω for $V_{DS}$ = 1 V and $V_{GS}$ = 50 V. From Ref. 16, the appearance of spin-valve effects requires the condition $R_{CH} < (R_{SB}/2) < R_{CH} \times (l_{sf}/t_{CH})^2$ (here, referred to as the conductivity matching condition), where $l_{sf}$ and $t_{CH}$ are the spin diffusion length in a channel semiconductor and the channel length, respectively. For the MnAs-MOSFET, $t_{CH}$ = 110 μm, when it is defined by the length between the center of the MnAs S/D. Although $l_{sf}$ in a MOS inversion channel has not been clarified, $l_{sf}$ in an undoped Si single crystal was deduced to be more than 320 μm.[6] Using these $l_{sf}$ and $t_{CH}$ values, the conductivity matching condition is fulfilled for $V_{DS} \geq 50$ mV and $V_{GS} \geq 30$ V, whereas $V_{DS}$ = 50 mV and $V_{GS} < 30$ V is not the case.

In order to characterize the spin-dependent transport, magnetoresistance for the MnAs MOSFET and the asymmetric MOSFETs was measured with constant $V_{DS}$ and $V_{GS}$ biases at 3 K, which was plotted by the output current $I_{DS}$ as a function of magnetic field. An in-plane external magnetic field $H$ was applied normal to the current flow. Hereafter, magnetoresistance is represented as the magneto-current ratio $\Gamma_{MC} = [I_{DS}(H) - I_{DS\_MIN}] / I_{DS\_MIN}$, where $I_{DS\_MIN}$ is the minimum current in the measurement. Figure 3(b) shows the $\Gamma_{MC}$ measured at 3 K for an asymmetric MOSFET with a 100-nm-thick MnAs film and another asymmetric MOSFET with a 20-nm-thick MnAs film when the $V_{DS}$ and $V_{GS}$ biases were 1 V and 50 V, respectively. By changing the polarity of the sweep, the $\Gamma_{MC}$ exhibits almost the symmetric behavior with respect to $\varGamma_{MC}$ axis for both cases. The external field, at which the $\varGamma_{MC}$ shows the minimum, reasonably agrees with the coercivities of the magnetization hysteresis loops measured at 3 K for a 100-nm-thick MnAs film and a 20-nm-thick MnAs film fabricated into 100 μm × 100 μm square patterns, as shown in Fig. 3(a). The linear slope of the $\varGamma_{MC}$ seen in the higher and lower fields than the minimum peaks in Fig. 3(b) originates from the forced effect, which was observed in MnAs films grown on GaAs(001) substrates.[17] Furthermore, the shape of $\varGamma_{MC}$ was basically the same when the S/D were alternated. Therefore, the $\varGamma_{MC}$ characteristics of Fig. 3(b) can be attributed to the magnetization-related resistance change originating from the MnAs contact itself. Note that the magnetization hysteresis loops in Fig. 3(a) are the total magnetic moment of a few tens of thousands square patterns. Thus, the individual hysteresis loop would have some deviations.

Figure 3(c) shows the $\varGamma_{MC}$ measured at 3 K for the MnAs MOSFET when the $V_{DS}$ and $V_{GS}$ biases were 1 V and 50 V, respectively. Almost the same hysteresis loop was reproducibly obtained. The most notable differences in $\varGamma_{MC}$ between the MnAs MOSFET and the asymmetric MOSFETs are the steep decrease at ±1.3 kOe. Compared with the $\varGamma_{MC}$ of the asymmetric MOSFETs, the features obtained for the MnAs MOSFET can not be explained by the superposition of the magnetization-related resistance change, such as AMR and LHE, originating from each of the MnAs S/D contacts. Moreover, the unique phenomena in a 2DEG channel at low temperatures, such as Anderson localization[18] and Rashba effect,[19] are excluded from the origin because of the in-plane magnetoresistance and the weak spin-orbit interaction in Si. Since the relative magnetization configuration between the MnAs S/D contacts is considered to be antiparallel in the range from 1.3 to 3 (and -1.3 to -3) kOe, we conclude that the MnAs MOSFET exhibits the spin-valve effect, that is, spin polarized electrons are injected from the MnAs source to the Si MOS inversion channel, and then transported through the 10-μm-long channel, and finally detected by the MnAs drain.

The $\Gamma_{MC}$ value decreased with increasing temperature, and the clear spin-valve-type hysteresis was obtained up to 50 K. This temperature dependence is also the clear evidence of the spin-valve effect since the magnitude of the magnetization, which determines the magnitude of AMR and LHE, does not change in this temperature range.[15] The magnetic fields of ±1.3 kOe and ±3.0 kOe, where the large $\Gamma_{MC}$ change is observed, are not exactly the same as the coercivities seen in the MnAs hysteresis loops. This is probably because the magnetization hysteresis loops of the S/D MnAs are somewhat different from those in Fig. 3(a) and/or partial magnetic domains mainly contribute to the spin-dependent transport.

The maximum $\Gamma_{MC}$ difference defined at around 1.3 kOe (referred to as $\Gamma_{MC\_MAX}$) is plotted as a function of $V_{DS}$ when the measurements were performed with $V_{GS}$ = 50 V, as shown in Fig. 4. Clear spin-valve characteristics as Fig. 3(c) was obtained for $V_{DS}$ above 50 mV, and the $\Gamma_{MC\_MAX}$ is almost independent of $V_{DS}$. On the other hand, almost same feature was obtained for $V_{DS}$ = 50 mV and $V_{GS}$ = 80 V, however, that feature was not obtained for $V_{DS}$ = 50 mV and $V_{GS}$ = 20 V. As a result, the clear spin-valve characteristics appear when the $R_{CH}$ satisfies the conductivity matching condition $R_{CH} < (R_{SB}/2) < R_{CH} \times (l_{sf}/t_{CH})^2$, which is consistent with the theory in Ref. 16. Since the charge transport in the MOS inversion channels is diffusive, the $\Gamma_{MC\_MAX}$ is probably determined by the spin injection/detection efficiency at the source-channel and channel-drain junctions if the $l_{sf}$ is far longer than the channel length. This consideration is consistent with the feature in Fig. 4, that is, the $\Gamma_{MC\_MAX}$ is insensitive to the $V_{DS}$ bias examined here.

The spin-valve effect will be also influenced by the spin polarization of the ferromagnetic electrodes. Whereas the spin polarization of type-II MnAs films grown on Si(001) is unknown, the epitaxial growth plane is the same with that of type-B MnAs films grown on GaAs(001). From the saturation magnetization, type-II MnAs films have a spin polarization considerably less than ~0.44 of type-B MnAs.[20] This can also lead to the small $\varGamma_{MC}$, and further investigation is needed to improve the $\varGamma_{MC}$ ratio.

In summary, spin-dependent transport was investigated in the spin MOSFET with ferromagnetic MnAs source and drain (S/D) contacts. In-plane magnetoresistance was measured with constant $V_{DS}$ and $V_{GS}$ biases for the spin MOSFET as well as for the MOSFETs with MnAs and nonmagnetic AuSb S/D contacts. By comparing the results for these MOSFETs, we conclude that the spin MOSFET exhibit the spin-valve effect originating from the spin-dependent transport through the Si channel.

This work was partly supported by Grant-in-Aids for Scientific Research, the Special Coordination Programs for Promoting Science and Technology, and R&D for Next-generation Information Technology by MEXT.

References

1. I. Žutić, J. Fabian, and S. D. Sarma, Rev. Mod. Phys. **76** (2004) 323.
2. SIA Semiconductor Industry Association, "The International Technology Roadmap for Semiconductors", San Jose, CA, 2005. http://public.itrs.net/
3. X. Lou, C. Adelmann, M. Furis, S. A. Crooker, C. J. Palmstrøm, and P. A. Crowell, Phys. Rev. Lett. **96** (2006) 176603; X. Lou, C. Adelmann, S. A. Crooker, E. S. Garlid, J. Zhang, K. S. Madhukar Reddy, S. D. Flexner, C. J. Palmstrøm, and P. A. Crowell, Nature Phys. **3** (2007) 197.
4. I. Appelbaum, B. Huang, and D.J. Monsma, Nature **447** (2007) 295.
5. B. Huang, D.J. Monsma, and I. Appelbaum, Phys. Rev. Lett. **99** (2007) 177209.
6. B. T. Jonker, G. Kioseoglou, A. T. Hanbicki, C. H. Li, and P. E. Thompson, Nature Phys. **3** (2007) 542.
7. O. M. J. van't Erve, A. T. Hanbicki, M. Holub, C. H. Li, C. Awo-Affouda, P. E. Thompson, and B. T. Jonker, Appl. Phys. Lett. **91** (2007) 212109.
8. T. Sasaki, T. Oikawa, T. Suzuki, M. Shiraishi, Y. Suzuki, and K. Tagami, Appl. Phys. Express **2** (2009) 053003.
9. S. Sugahara and M. Tanaka, Appl. Phys. Lett. **84**, 2307 (2004); J. Appl. Phys. **97** (2005) 10D503; ACM Transactions on Strage **2** (2006) 197.
10. S. Sugahara, IEE Proc.-Circuits, Devices-Sys. **152** (2005) 355.
11. M. Tanaka and S. Sugahara, IEEE Trans. Electron Devices **54** (2007) 961.
12. F. G. Monzon and M. L. Roukes, J. Magn. Magn. Matr. **198** (1999) 628; F. G. Monzon, D. S. Patterson, and M. L. Roukes, J. Magn. Magn. Mater. **195** (1999) 19.
13. K. Akeura, M. Tanaka, M. Ueki, and T. Nishinaga, Appl. Phys. Lett. **67** (1995) 1.
14. K. Sugiura, R. Nakane, S. Sugahara, and M. Tanaka, Appl. Phys. Lett. **89** (2006) 072110.
15. K. Sugiura, R. Nakane, S. Sugahara, and M. Tanaka, J. Cryst. Growth **301–302** (2007) 611.
16. A. Fert and H. Jaffrès, Phys. Rev. B **64** (2001) 184420.
17. R. Nakane, S. Sugahara, and M. Tanaka, J. Appl. Phys. **95** (2004) 6558.
18. Y. Kawaguchi and S. Kawaji, Surf. Sci. **73** (1978) 520.
19. Y. A. Bychkov and E. I. Rashba, J. Phys. C **17** (1984) 6039.
20. R. P. Panguluri, G. Tsoi1, B. Nadgorny, S. H. Chun, N. Samarth, and I. I. Mazin, Phys. Rev. B **68** (2003) 201307.

Figure captions

Figure1 (a) Schematic device structure and (b) optical micrograph (top view) of a fabricated spin MOSFET with ferromagnetic MnAs source and drain (S/D) contacts, where the buried oxide layer and the bottom Si substrate were used as a gate dielectric and a gate electrode, respectively.

Figure 2(a) Source-drain current $I_{DS}$ vs. source-drain bias voltage $V_{DS}$ output characteristics measured at 3 K for a MnAs MOSFET with $L_G$ = 10 μm, where the source−gate bias $V_{GS}$ varied from 0 to 100 V in steps of 10 V.  (b) $I_{DS}$ vs. $V_{GS}$ output characteristics measured with $V_{DS}$ = 50 mV at 3 K for the same MnAs-MOSFET in (a) and a n-MOSFET with $L_G$ = 10 μm.

Figure 3(a)  Magnetization hysteresis loops at 3 K of a 100-nm-thick and a 20-nm-thick MnAs films fabricated into 100 μm × 100 μm square patterns.  (b)  Magneto-current ratio $\Gamma_{MC}$ measured at 3 K for an asymmetric MOSFET with a 100-nm-thick MnAs film and another asymmetric MOSFET with a 20-nm-thick MnAs film when the $V_{DS}$ and $V_{GS}$ biases were 1 V and 50 V, respectively.  Here, the magneto-current ratio is defined by $\Gamma_{MC}$ = [$I_{DS}(H)$ − $I_{DS\_MIN}$] / $I_{DS\_MIN}$, where $I_{DS\_MIN}$ is the minimum current in a measurement.  (c) $\Gamma_{MC}$ measured at 3 K for the MnAs MOSFET when the $V_{DS}$ and $V_{GS}$ biases were 1 V and 50 V, respectively.  The red and blue arrows represent the positive (from 10 to -10 kOe) and the negative (from -10 to 10 kOe) sweep of the external field, respectively.  The open and closed arrows represent the magnetization directions of the 20-nm-thick MnAs and the 100-nm-thick MnAs, respectively.

Figure 4  $V_{DS}$ bias dependence of $\Gamma_{MC\_MAX}$ measured for the MnAs MOSFET at 3 K when the $V_{GS}$ bias was 50 V.

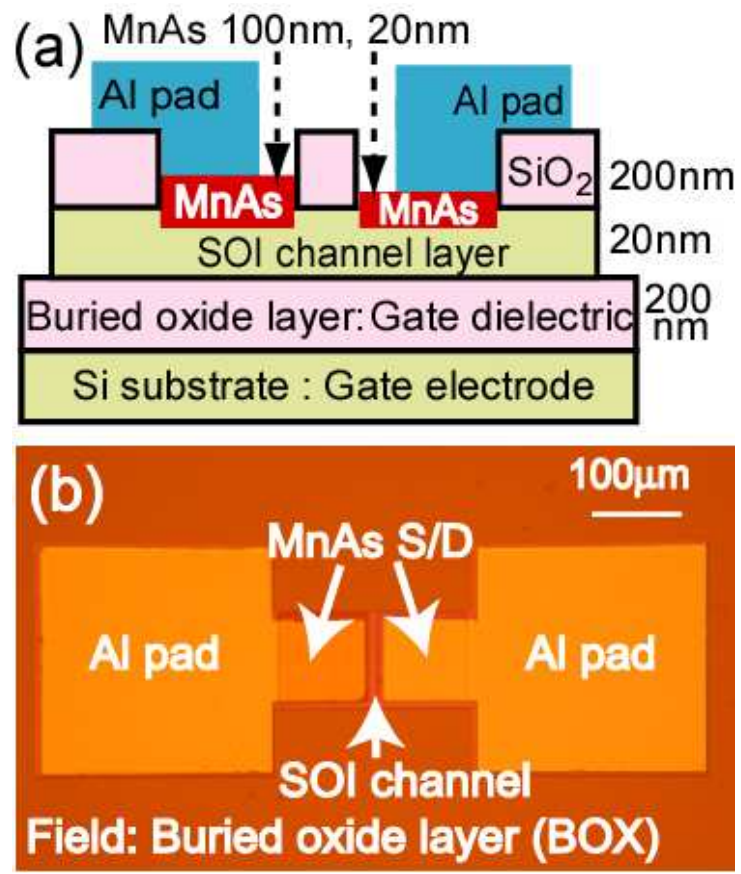

Nakane et al.　Fig. 1

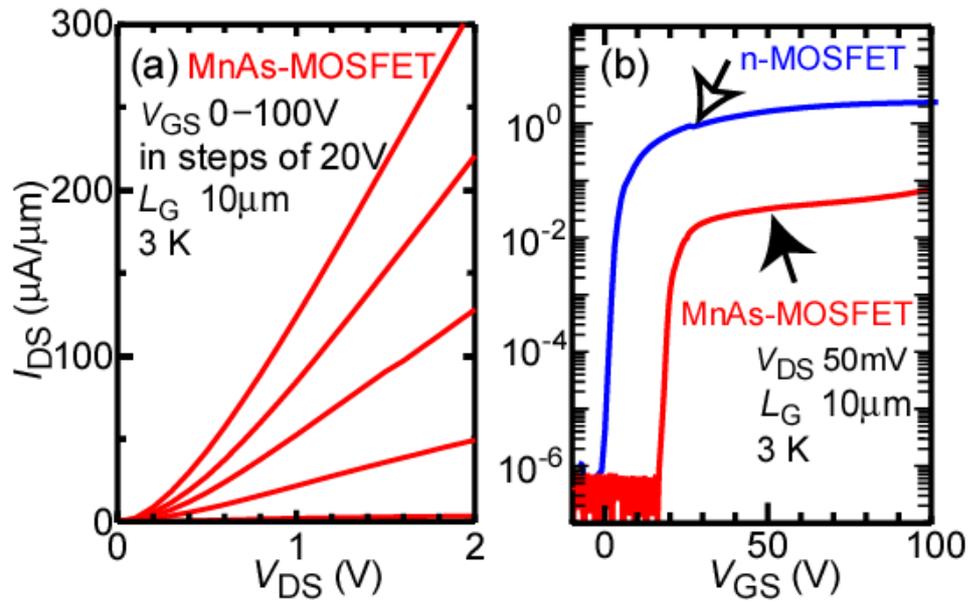

Nakane et al.   Fig. 2

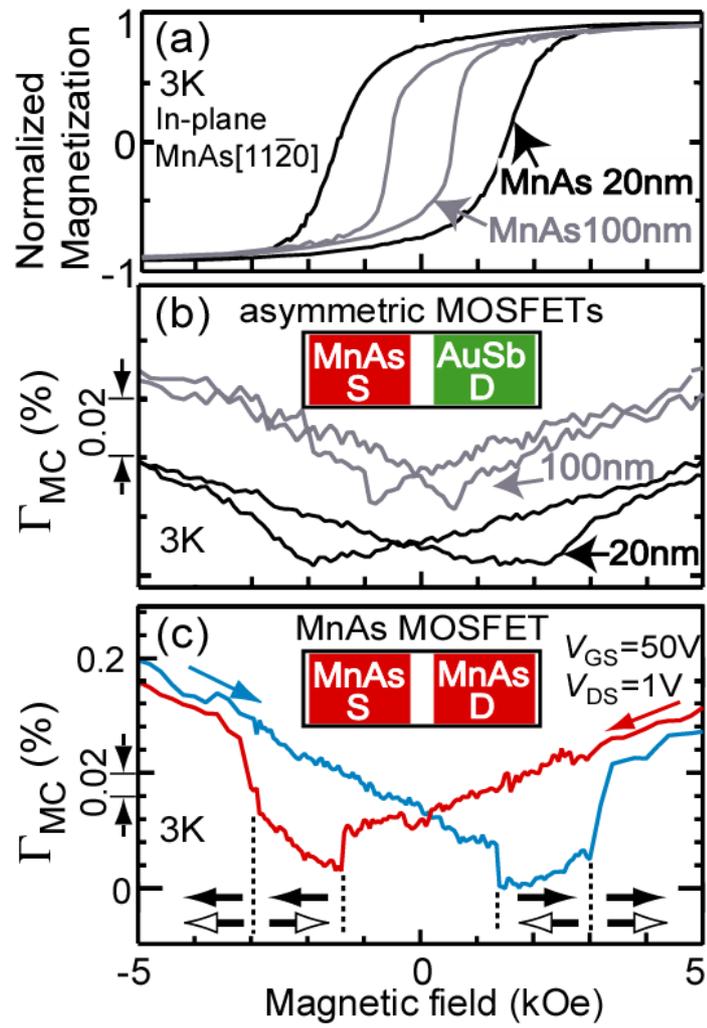

Nakane et al.   Fig. 3

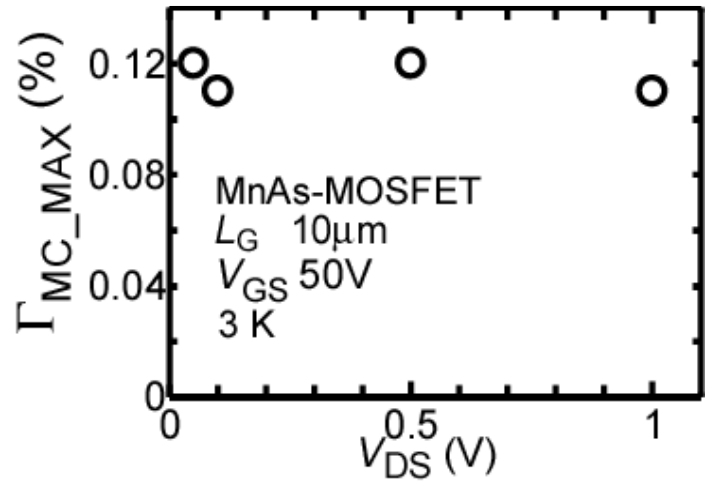

Nakane et al.    Fig. 4